\documentclass[twocolumn]{aastex62}

\shorttitle{{\sl Gaia} EDR3 Parallax Zero-point Offset based on Contact Binaries}
\shortauthors{F.-Z. Ren et al.}

\begin{document}

\title{{\sl Gaia} EDR3 Parallax Zero-point Offset based on W Ursae
  Majoris-type Eclipsing Binaries}

\correspondingauthor{Fangzhou Ren, Xiaodian Chen}
\email{renfz@pku.edu.cn \\ chenxiaodian@nao.cas.cn}

\author[0000-0003-4623-0031]{Fangzhou Ren}
\affiliation{Department of Astronomy, School of Physics, Peking
  University, Yi He Yuan Lu 5, Hai Dian District, Beijing 100871,
  People's Republic of China}
\affiliation{Kavli Institute for Astronomy and Astrophysics, Peking
  University, Yi He Yuan Lu 5, Hai Dian District, Beijing 100871,
  People's Republic of China}

\author[0000-0001-7084-0484]{Xiaodian Chen}
\affiliation{CAS Key Laboratory of Optical Astronomy, National
  Astronomical Observatories, Chinese Academy of Sciences, Beijing
  100101, People's Republic of China}
\affiliation{School of Astronomy and Space Science, University of the
  Chinese Academy of Sciences, Beijing 101408, China}
\affiliation{Department of Astronomy, China West Normal University,
  Nanchong, 637009, China}

\author[0000-0002-7727-1699]{Huawei Zhang}
\affiliation{Department of Astronomy, School of Physics, Peking
  University, Yi He Yuan Lu 5, Hai Dian District, Beijing 100871,
  People's Republic of China}
\affiliation{Kavli Institute for Astronomy and Astrophysics, Peking
  University, Yi He Yuan Lu 5, Hai Dian District, Beijing 100871,
  People's Republic of China}

\author[0000-0002-7203-5996]{Richard de Grijs}
\affiliation{Department of Physics and Astronomy, Macquarie
  University, Balaclava Road, Sydney, NSW 2109, Australia}
\affiliation{Research Centre for Astronomy, Astrophysics and
  Astrophotonics, Macquarie University, Balaclava Road, Sydney, NSW
  2109, Australia}

\author[0000-0001-9073-9914]{Licai Deng}
\affiliation{CAS Key Laboratory of Optical Astronomy, National
  Astronomical Observatories, Chinese Academy of Sciences, Beijing
  100101, People's Republic of China}
\affiliation{School of Astronomy and Space Science, University of the
  Chinese Academy of Sciences, Beijing 101408, China}
\affiliation{Department of Astronomy, China West Normal University,
  Nanchong, 637009, China}

\author{Yang Huang}
\affiliation{South-Western Institute for Astronomy Research, 
Yunnan University, Kunming 650500, People's Republic of China}

\begin{abstract}
We independently determine the zero-point offset of the {\sl Gaia}
early Data Release-3 (EDR3) parallaxes based on $\sim 110,000$ W Ursae
Majoris (EW)-type eclipsing binary systems. EWs cover almost the
entire sky and are characterized by a relatively complete coverage in
magnitude and color. They are an excellent proxy for Galactic
main-sequence stars. We derive a $W1$-band Period-Luminosity relation
with a distance accuracy of 7.4\%, which we use to anchor the {\sl
  Gaia} parallax zero-point. The final, global parallax offsets are
$-28.6\pm0.6$ $\mu$as and $-25.4\pm4.0$ $\mu$as (before correction)
and $4.2\pm0.5$ $\mu$as and $4.6\pm3.7$ $\mu$as (after correction) for
the five- and six-parameter solutions, respectively. The total
systematic uncertainty is $1.8$ $\mu$as. The spatial distribution of
the parallax offsets shows that the bias in the corrected {\sl Gaia}
EDR3 parallaxes is less than 10 $\mu$as across 40\% of the sky. Only
15\% of the sky is characterized by a parallax offset greater than 30
$\mu$as. Thus, we have provided independent evidence that the parallax
zero-point correction provided by the {\sl Gaia} team significantly
reduces the prevailing bias. Combined with literature data, we find
that the overall {\sl Gaia} EDR3 parallax offsets for Galactic stars
are $[-20, -30]$ $\mu$as and 4--10 $\mu$as, respectively, before and
after correction. For specific regions, an additional deviation of
about 10 $\mu$as is found.
\end{abstract}

\section{Introduction} \label{sec:intro}

{\sl Gaia}'s early third data release
\citep[EDR3;][]{2020arXiv201201533G} includes astrometric and
photometric measurements of more than 1.81 billion sources brighter
than $G = 21$ mag. More than 1.46 billion have triangulated parallax
measurements with typical uncertainties of 0.03--1.3 mas for stars
with $15 < G < 21$ mag. Although the {\sl Gaia} mission represents a
leap forward for tests of stellar and Galactic astrophysics,
systematic parallax errors are inevitably caused by imperfections in
the instruments and data processing \citep{2020arXiv201203380L}. For
small parallaxes, the effects of systematic errors are significant,
which thus calls for examination of the {\sl Gaia} EDR3 parallaxes
using a variety of independent distance tracers.

Comparison of the {\sl Gaia} catalog with other compilations is
essential for evaluation of the data quality, and thus to understand
{\sl Gaia}'s performance. Special attention has been paid to detecting
possible biases. A parallax zero-point offset was noted from the first
{\sl Gaia} data release
\citep{2016A&A...595A...1G,2016A&A...595A...4L}; it has since been
confirmed
\citep[e.g.][]{2016ApJ...832L..18J,2016ApJ...831L...6S}. This offset
persisted in {\sl Gaia} DR2 \citep{2018A&A...616A...1G}, ranging from
$-29$ $\mu$as to $-80$ $\mu$as
\citep[e.g.][]{2019ApJ...878..136Z}. {\sl Gaia} EDR3 represents a
significant improvement with respect to {\sl Gaia} DR2
\citep{2020arXiv201206242F} as regards the resulting parallaxes,
astrometric parameters, and parallax zero-point corrections.

\citet{2020arXiv201201742L} published a formal procedure to remove the
parallax zero-point offset, which is a function of stellar magnitude,
color, and spatial position. Their correction was based on quasars
distributed across the entire sky, stars in the Large Magellanic
Cloud, and physical binaries. The calibration models differ for
astrometric solutions with either five or six parameters. The
corrections are most appropriate for sources with similar magnitudes
and colors as those in the quasar sample (faint and blue) rather than
for typical Galactic stars. Therefore, independent validation of {\sl
  Gaia} parallaxes based on Galactic objects is urgently needed.

Eclipsing binary systems (EBS) exhibit optical variability because of
geometric properties rather than due to intrinsic physical
variations. Both components of W Ursae Majoris (EW)-type EBS fill
their Roche lobes and have similar temperatures. The primary component
is similar to a main-sequence star. EWs can be used as distance
indicators, because they follow a well-defined period--luminosity
relation
\citep[PLR;][]{1997PASP..109.1340R,2018ApJ...859..140C}. Particularly in
infrared (IR) bands, a single EW system can yield a distance with 8\%
accuracy. In recent decades, the number of known EWs has grown
exponentially thanks to new, large surveys \citep[][and references
  therein]{2021AJ....161..176R}.

Here, we use EW distances to investigate the zero-point offset in {\sl
  Gaia} EDR3 parallaxes. EWs are among the most numerous variables in
the Milky Way for which distances can be determined
independently. Section 2 introduces our data set, Section 3 presents
our method and the main results, and Section 4 discusses our
systematic errors and a comparison with literature results. Section 5
summarizes our conclusions.

\section{Data} \label{sec:data}

We used EW data from the American Association of Variable Star
Observers International Variable Star
Index\footnote{\url{https://www.aavso.org/vsx/index.php}}
\citep{2006SASS...25...47W}. This catalog contains 0.4 million EWs,
most of which come from the Zwicky Transient Facility's
\citep[ZTF;][]{2020ApJS..249...18C} variables catalog. The remainder
originate from the All-Sky Automated Survey for Supernovae
\citep[ASAS-SN;][]{2018MNRAS.477.3145J}, the Asteroid
Terrestrial-impact Last Alert System
\citep[ATLAS;][]{2018AJ....156..241H}, and the Wide-field Infrared
Survey Explorer ({\sl WISE}) catalog of periodic variable stars
\citep{2018ApJS..237...28C}.

We selected EWs with periods of $-0.55<\log P \mbox{ (days)}<-0.25$,
which follow a tight PLR, comprising 144,777 objects in {\sl Gaia}
EDR3. These EBS were cross-matched with the {\sl WISE} database to
obtain $W1$ amplitudes. We used the $W1$ band to determine their
distances, because the extinction in this IR band is much smaller than
in optical bands and also since $W1$ magnitudes are average EW
magnitudes based on $\sim 30$ independent detections. More
importantly, the $W1$ PLR is the most accurate EW PLR \citep[][their
  Fig. 2]{2018ApJ...859..140C}. For data quality control, we applied
as additional criteria:

\begin{enumerate}

\item Renormalized unit weight error (RUWE)\,$\leq 1.4$;

\item Blending factor \,$\leq 1.1$.

\end{enumerate}

The RUWE is equivalent to an astrometric goodness-of-fit
indicator. Larger values indicate that the astrometric solution does
not completely describe the source motion
\citep{2020arXiv201203380L,2020arXiv201206242F};  this usually
  implies the presence of a tertiary companion and, hence, results in
  significant parallax differences
  \citep{2021ApJ...907L..33S}. Although we only consider EWs with
  RUWE\,$\leq$ 1.4, this will not significantly affect our results
  (the difference for the overall offset is less than 1$\mu$as). For a
  typical separation between EW components of $5 R_\odot$
  \citep{2021AJ....161..176R}, their angular semi-major axis is 0.0048
  mas. Since the photocenter's semi-major axis is clearly smaller than
  the angular semi-major axis, and because EWs are common-envelope
  objects with significantly reduced photocenter semi-major axes, the
  photocenter motion \citep{2021ApJ...907L..33S} is negligible in our
  sample. The blending factor is used to correct $W1$ magnitudes and
exclude EWs that are significantly affected by bright neighbors in the
$W1$ band, since the {\sl WISE} angular resolution ($6 \arcsec$ in
$W1$) is worse than that of {\sl Gaia} EDR3 \citep[about $2
  \arcsec$][their Fig. 6]{2020arXiv201203380L}. The blending factor is
the ratio of the total {\sl Gaia} $G$-band luminosity of all sources
within a $3 \arcsec$ radius around the target to the target
luminosity.

Application of our selection criteria resulted in, respectively,
109,512 and 4309 EWs with five- and six-parameter solutions in {\sl
  Gaia} EDR3. They are distributed across the full sky, except for
small regions near the Galactic Center and in the southern midplane.

\section{Results} \label{sec:result}

\subsection{PLRs for EWs} \label{sec:PLR}

\citet{2018ApJ...859..140C} determined optical--to--mid-IR PLRs based
on 183 EWs with {\sl Tycho--Gaia} parallaxes. We rederived the $W1$
PLR using {\sl Gaia} EDR3 parallaxes to improve the PLR
zero-point. Although only the maximum EW magnitudes, i.e., those
outside eclipses, are directly related to the periods, a tight
relationship also exists between mean magnitudes and periods; the
dispersion between maximum and mean magnitudes is just $\sigma=0.05$
mag \citep{2018ApJ...859..140C}. Our most important reason for
deriving the mean-magnitude PLR is that it is more appropriate and
convenient for large samples. Maximum magnitudes cannot be determined
easily, especially not for EWs collected from different catalogs.

\begin{figure}[ht!]
\centering
\includegraphics[width=8cm]{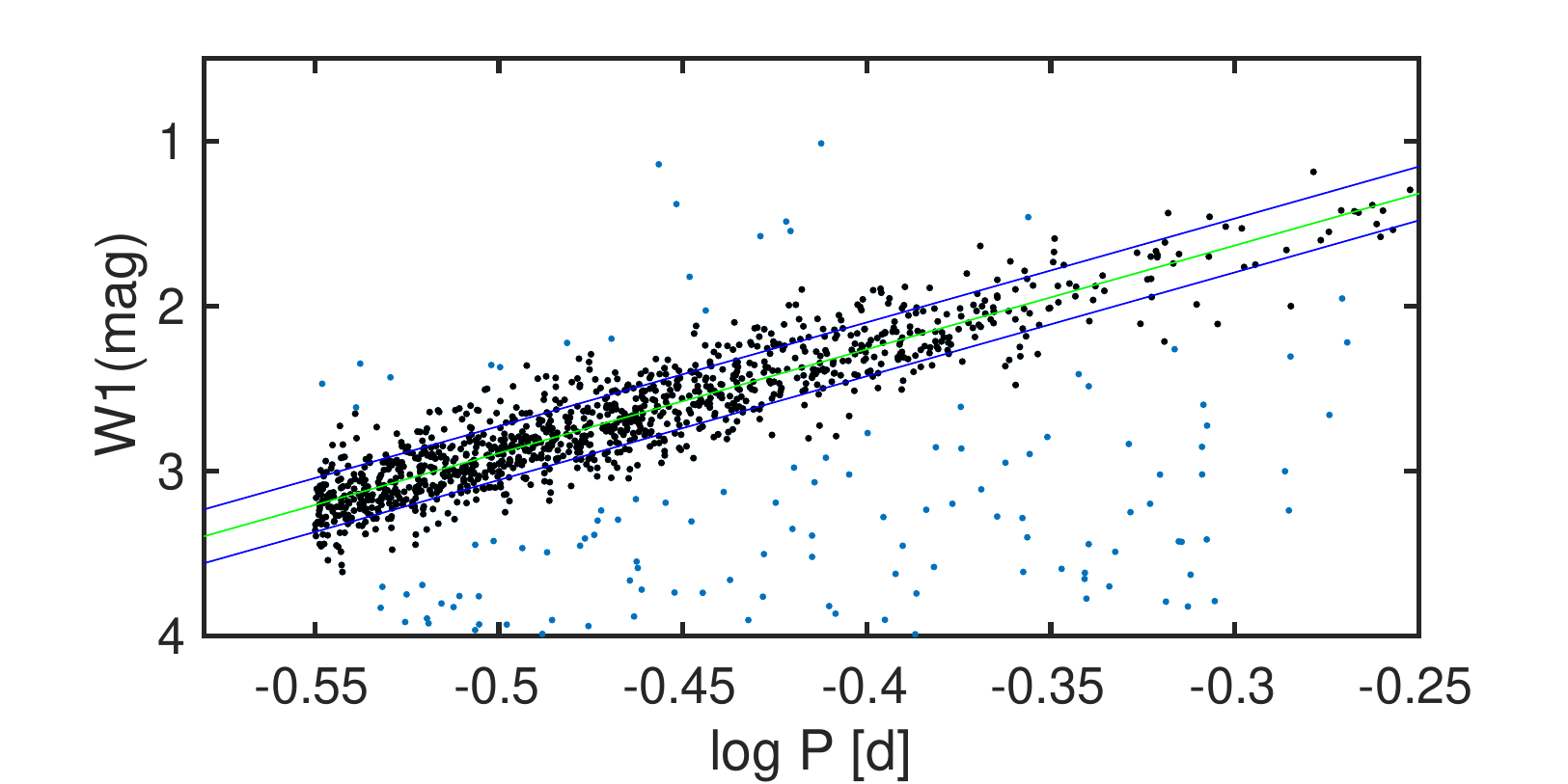}
\includegraphics[width=8cm]{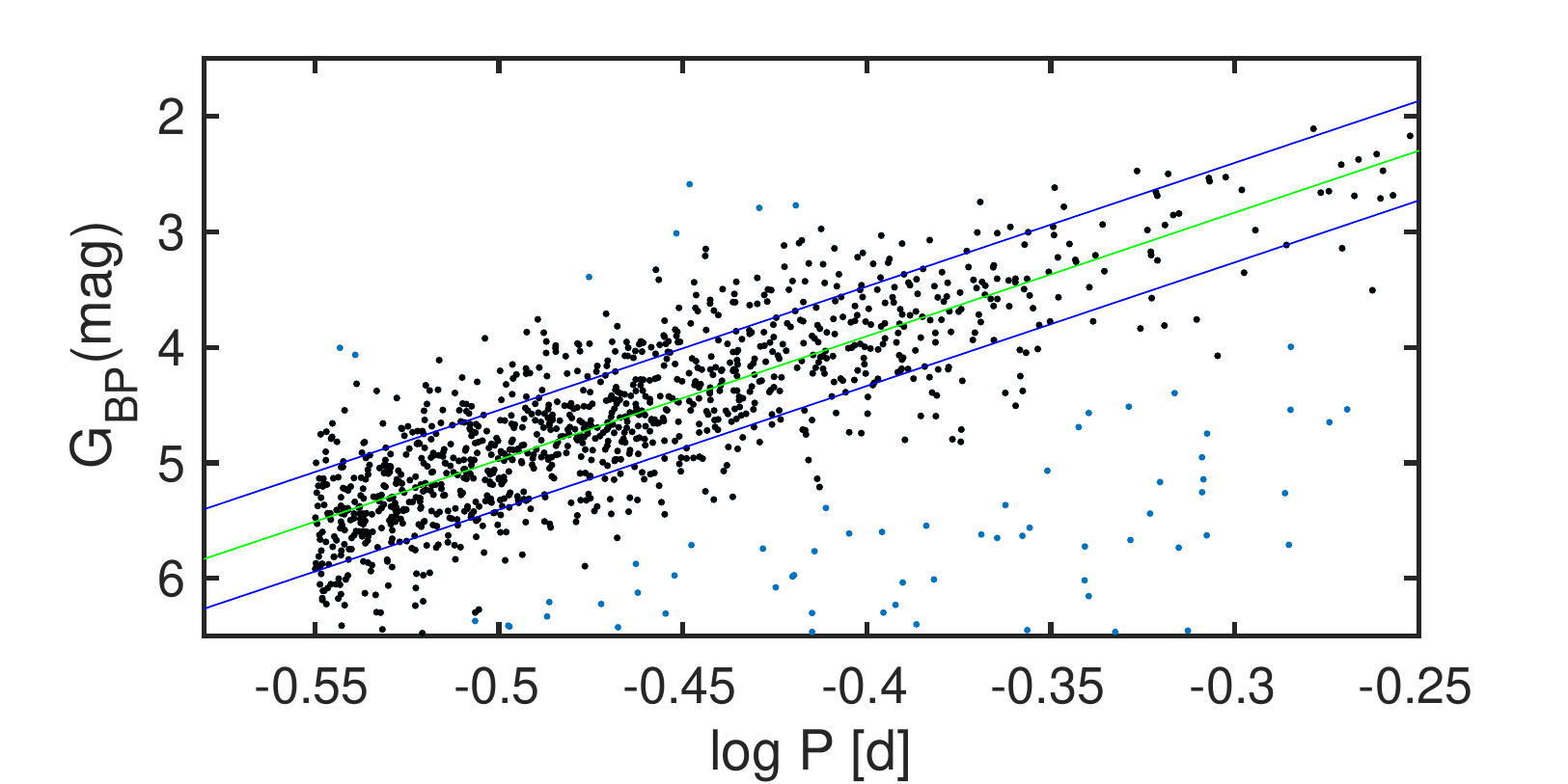}
\includegraphics[width=8cm]{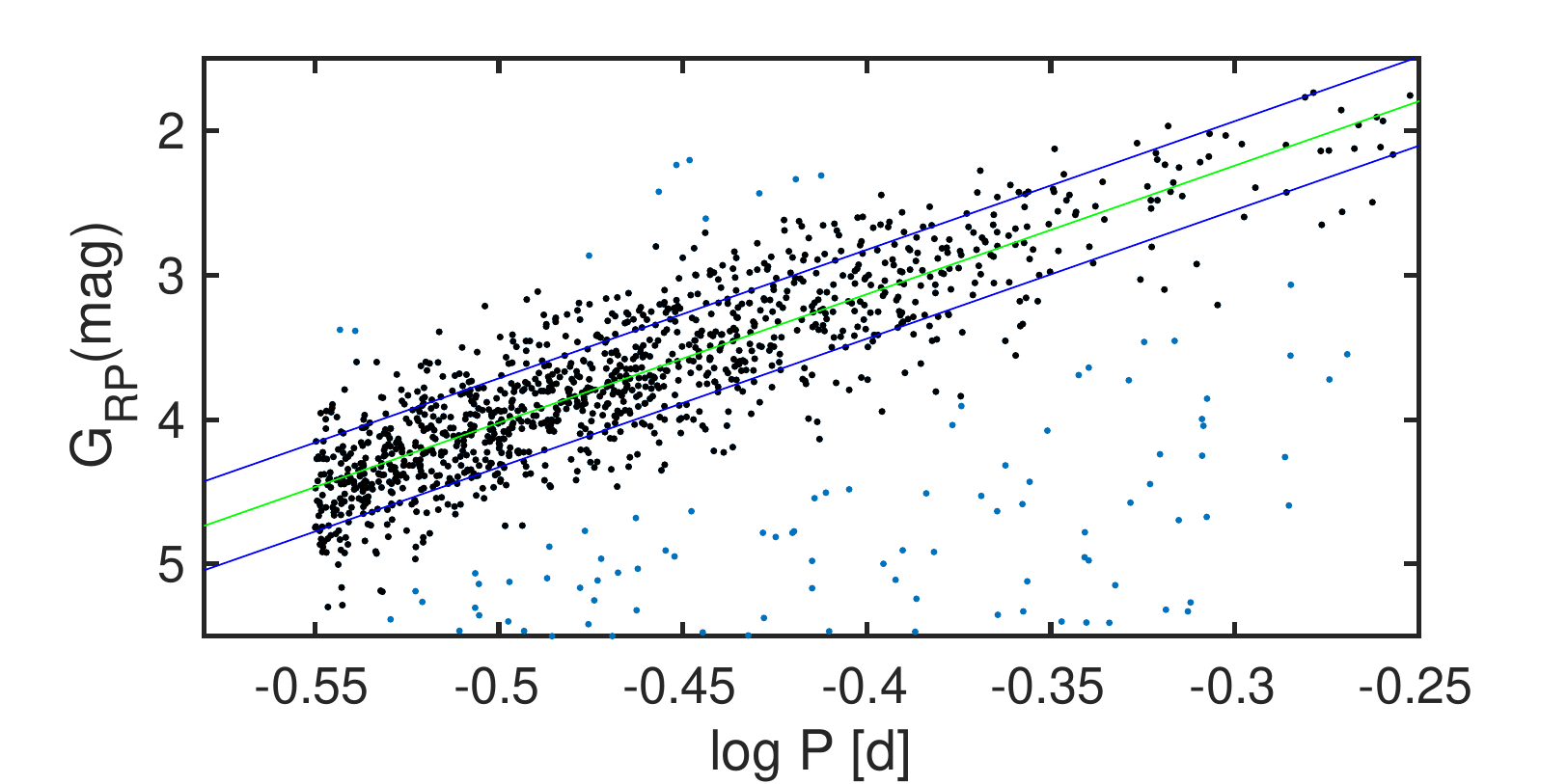}
\caption{(Top) $W1$, (middle) $G_{\rm BP}$, and (bottom) $G_{\rm RP}$
  PLRs for nearby EWs with accurate parallaxes. Black points: Objects
  adopted for the PLR determination. Blue dots: Outliers located
  outside the $3\sigma$ envelope. Green line: Best fit. Blue lines: $1
  \sigma$ range.
\label{fig:PLR}}
\end{figure}

We selected nearby ($< 500$ pc), bright EWs with accurate parallaxes
($\sigma_{\pi}/\pi < 0.01$, where $\pi$ and $\sigma_{\pi}$ are the
parallax and its uncertainty, respectively) and $W1$ magnitudes
($\sigma_{W1} < 0.05$ mag). We only consider nearby stars to
  reduce the systematic error, since the bias (zero-point offset
  versus parallax) is smaller for nearby stars. The systematic error
  associated with the zero-point offset in our PLR fits is
  proportionally reduced when applied to additional EWs (see Section
  \ref{sec:systematicerror} for further details). Extinction values
were estimated using the three-dimensional (3D) dust reddening map of
\citet{2019ApJ...887...93G} and $A_{W1}/A_V=0.039$
\citep{2019ApJ...877..116W}. We determined absolute magnitudes via
$M_{W1}=m_{W1}-5\log(1000/\pi)+5-A_{W1}$, where the unit of $\pi$ is
mas and $m_{W1}$ is the mean magnitude. The $W1$ PLR was determined
from a linear fit to the 1138 objects contained within the $3\sigma$
envelope (see Figure \ref{fig:PLR}, top), $M_{W1}=(-6.27\pm0.15)\log P
\mbox{ (days)} -0.24\pm0.07,\sigma=0.16$ mag. The green and blue lines
in Figure \ref{fig:PLR} indicate the linear fit and the $1 \sigma$
range, respectively.

The average extinction for our 1138 EWs is $A_{W1}=0.013$ mag
($A_V=0.322$ mag). This is reliable and appropriate for a sample at an
average distance of 368 pc. Considering a 10\% uncertainty in the
extinction, reflecting uncertainties due to our choice of extinction
law, the prevailing systematic bias is around 0.0013 mag.

The $G_{\rm BP}$ and $G_{\rm RP}$ PLRs were determined similarly (see
Figure \ref{fig:PLR}, middle and bottom): $M_{BP}=(-10.71\pm0.40) \log
P-0.38\pm0.19,\sigma=0.43$ mag and $M_{RP}=(-8.98\pm0.29) \log
P-0.44\pm0.13,\sigma=0.31$ mag. They were used for extinction
estimates in the part of the southern sky not covered by the 3D
extinction map (see Section \ref{sec:zero}).

\subsection{{\sl Gaia} EDR3 zero-point offset} \label{sec:zero}

\begin{figure}[ht!]
\centering
\includegraphics[width=8cm]{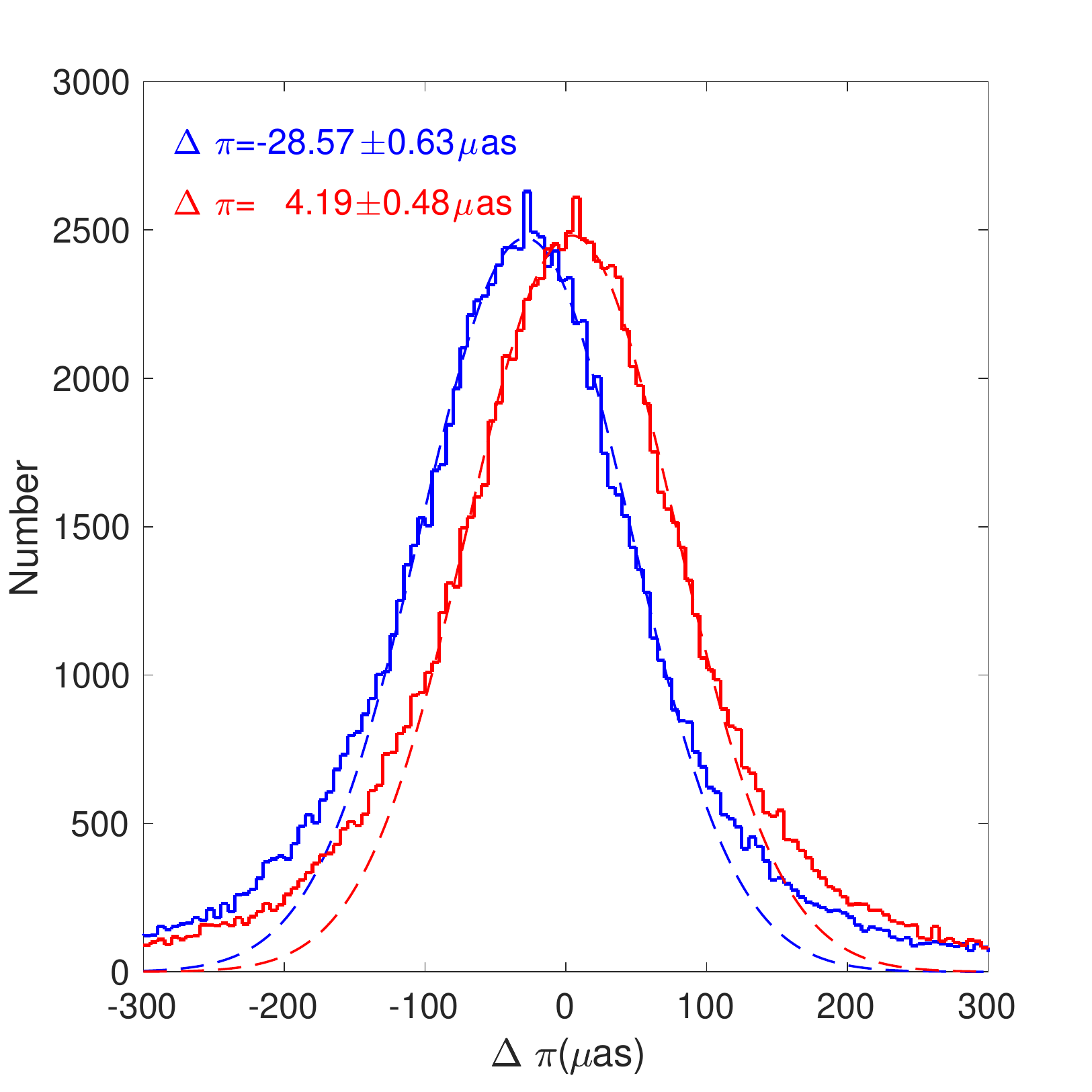}
\includegraphics[width=8cm]{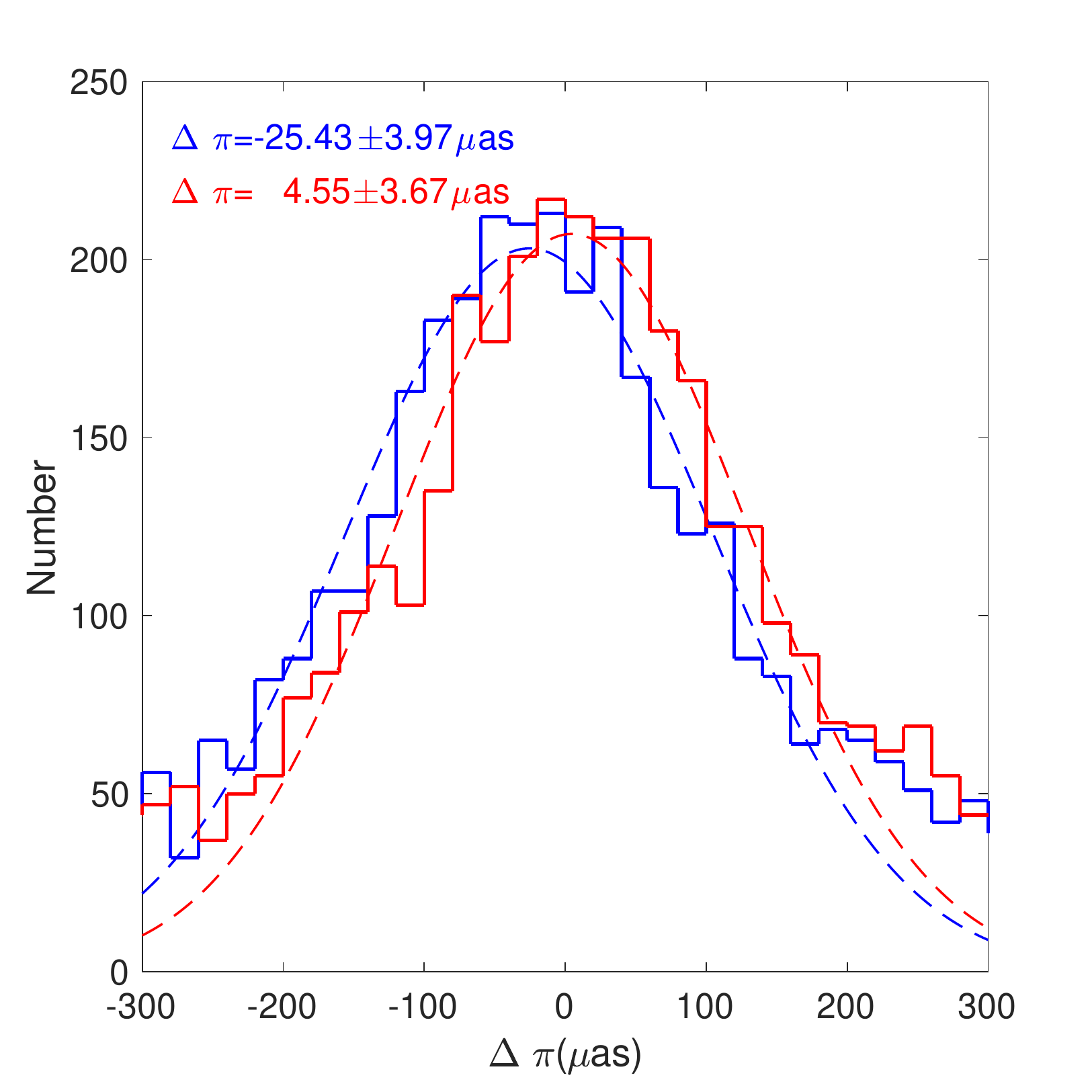}
\caption{Comparison of predicted parallaxes from the PLR versus {\sl
    Gaia} EDR3 parallaxes. (top) Objects with five-parameter
  solutions. The blue and red histograms represent the $(\pi_{\rm
    EDR3}-\pi_{\rm EW})$ and $(\pi_{\rm EDR3}^{\rm corr}-\pi_{\rm
    EW})$ distributions, respectively. The blue and red dashed lines
  represent Gaussian fits; the corresponding mean differences are
  indicated. (bottom) As the top panel, but for six-parameter
  solutions.\label{fig:zero}}
\end{figure}

\begin{figure*}
\begin{center}
\includegraphics[width=8cm]{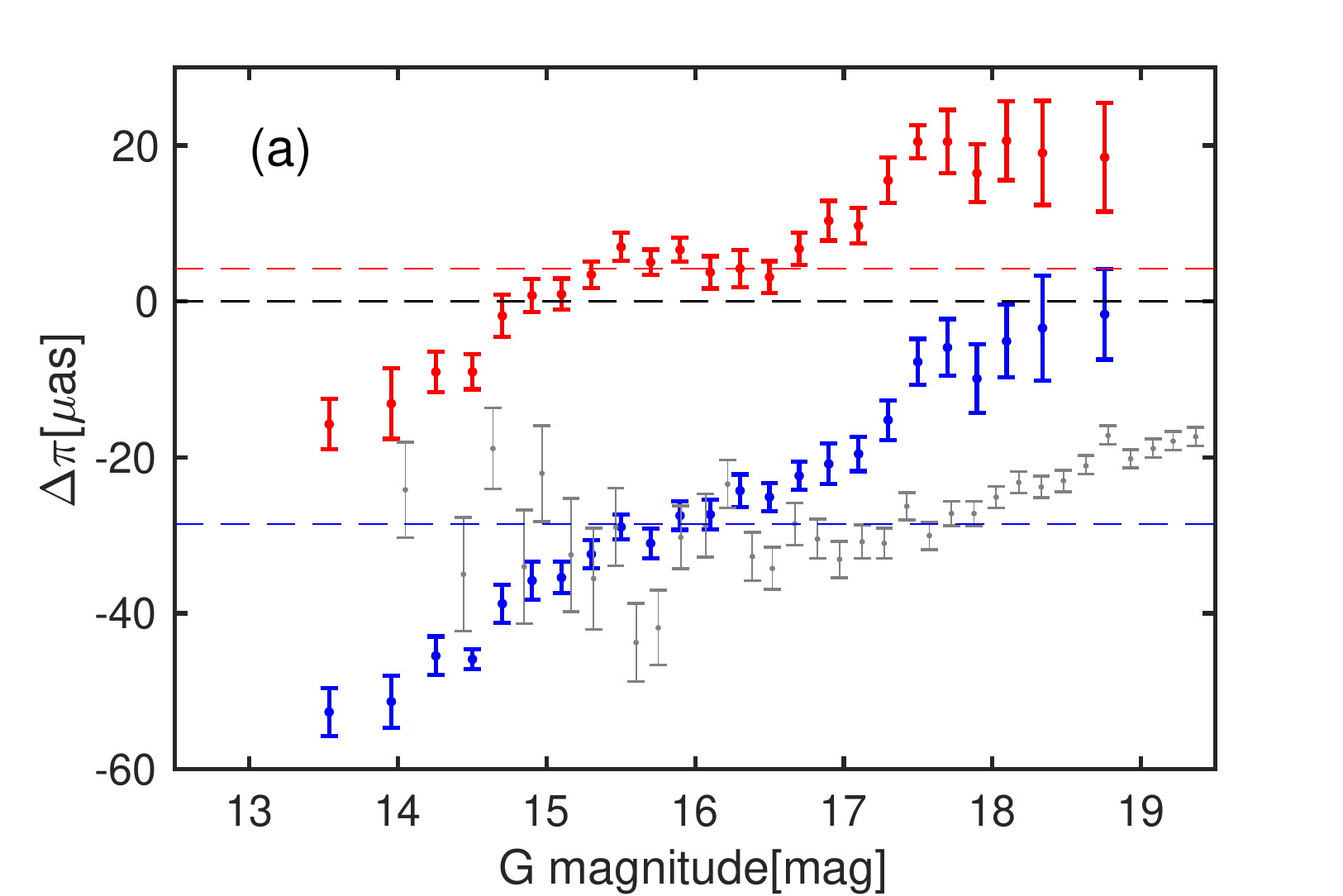}
\includegraphics[width=8cm]{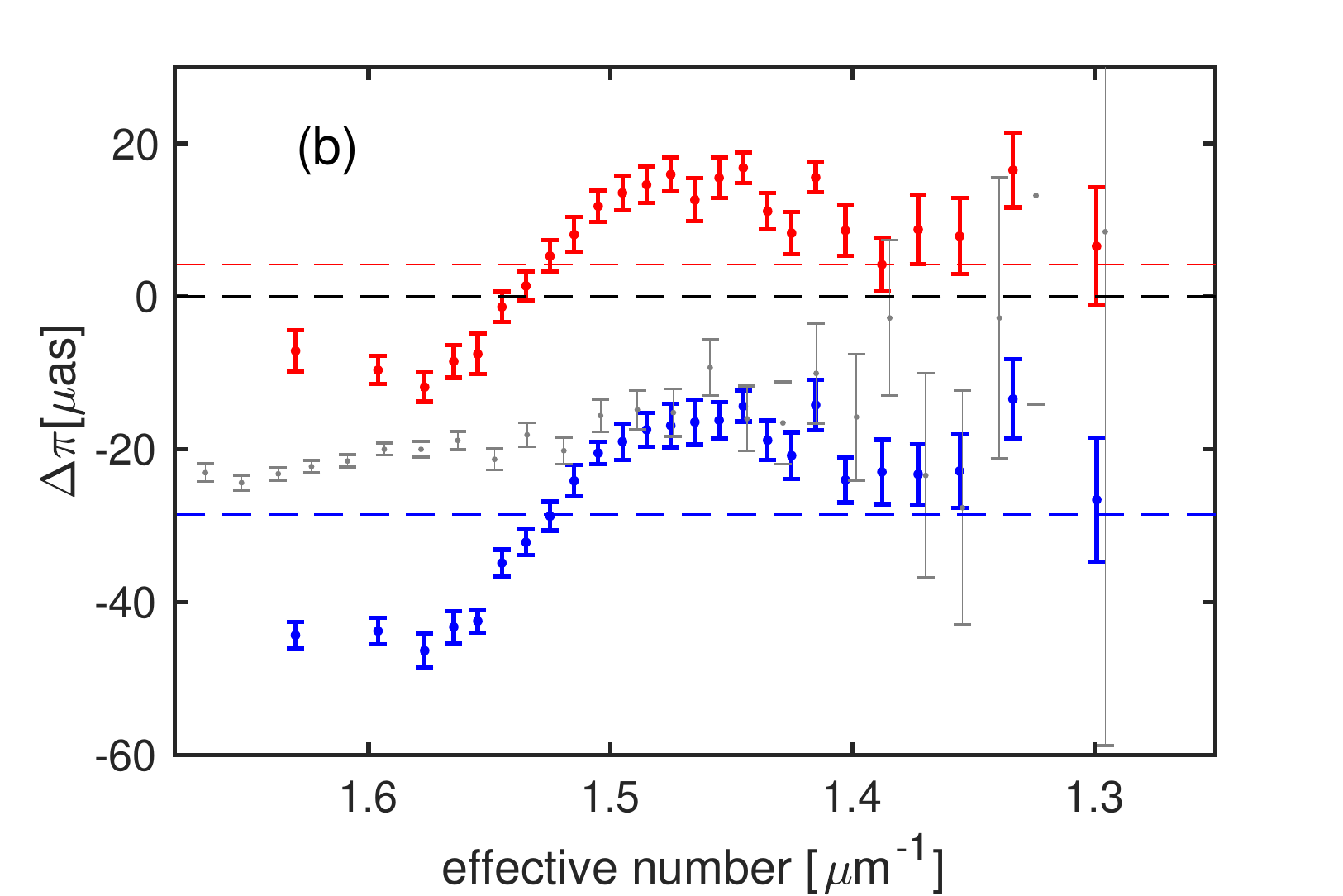}
\includegraphics[width=8cm]{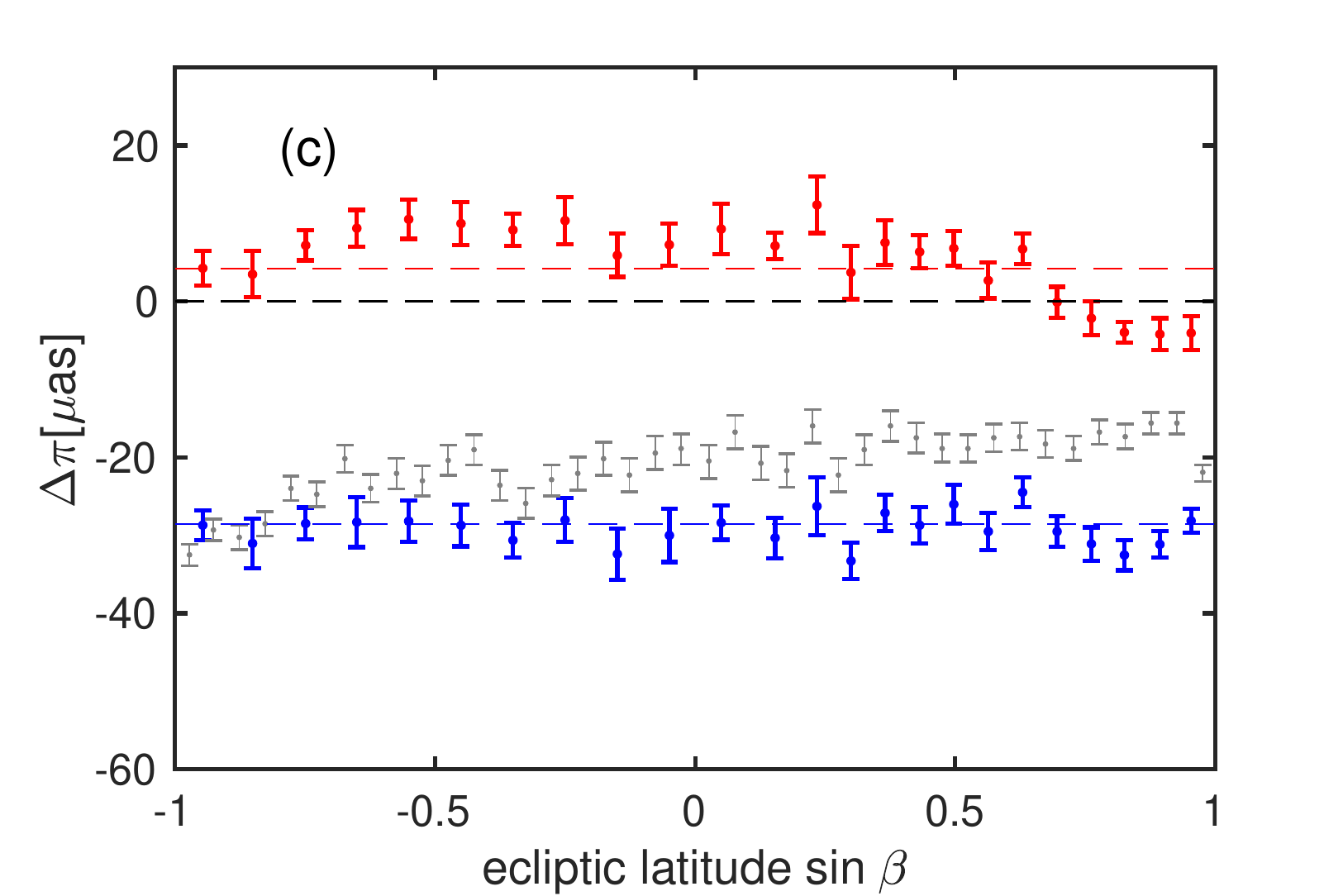}
\includegraphics[width=8cm]{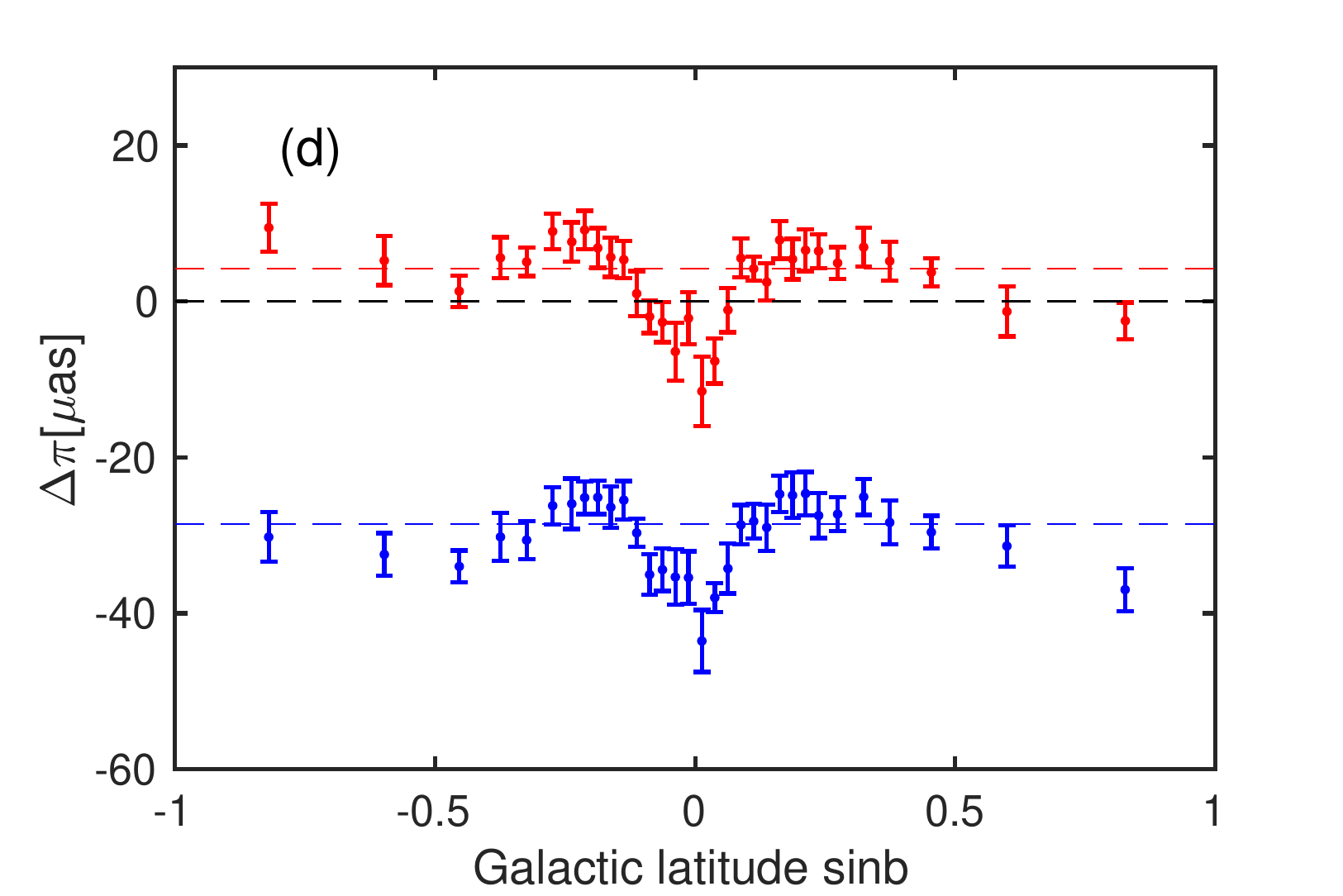}
\caption{Parallax differences for EWs between the {\sl Gaia} EDR3
  five-parameter solution and the PLR as a function of (a) $G$
  magnitude, (b) effective wavenumber, $\nu_{\rm eff}$, (c) ecliptic
  latitude, $\sin\beta$, (d) Galactic latitude, $\sin b$. The blue
  points and error bars represent the bias in $\Delta \pi_{\rm EDR3}$;
  the red symbols pertain to $\Delta \pi_{\rm EDR3}^{\rm corr}$,
  obtained from the best Gaussian fits and their standard
  deviations. The numbers of stars within each bin exceed 2000. Gray
  points relate to our quasar control sample \citep[][their
    Fig. 5]{2020arXiv201201742L}. The blue, red, and black dashed
  lines are the overall offsets in $\Delta \pi_{\rm EDR3}$ and $\Delta
  \pi_{\rm EDR3}^{\rm corr}$, and the zero deviations, respectively.
\label{fig:detail}}
\end{center}
\end{figure*}

\begin{figure}[ht!]
\centering
\includegraphics[width=8cm]{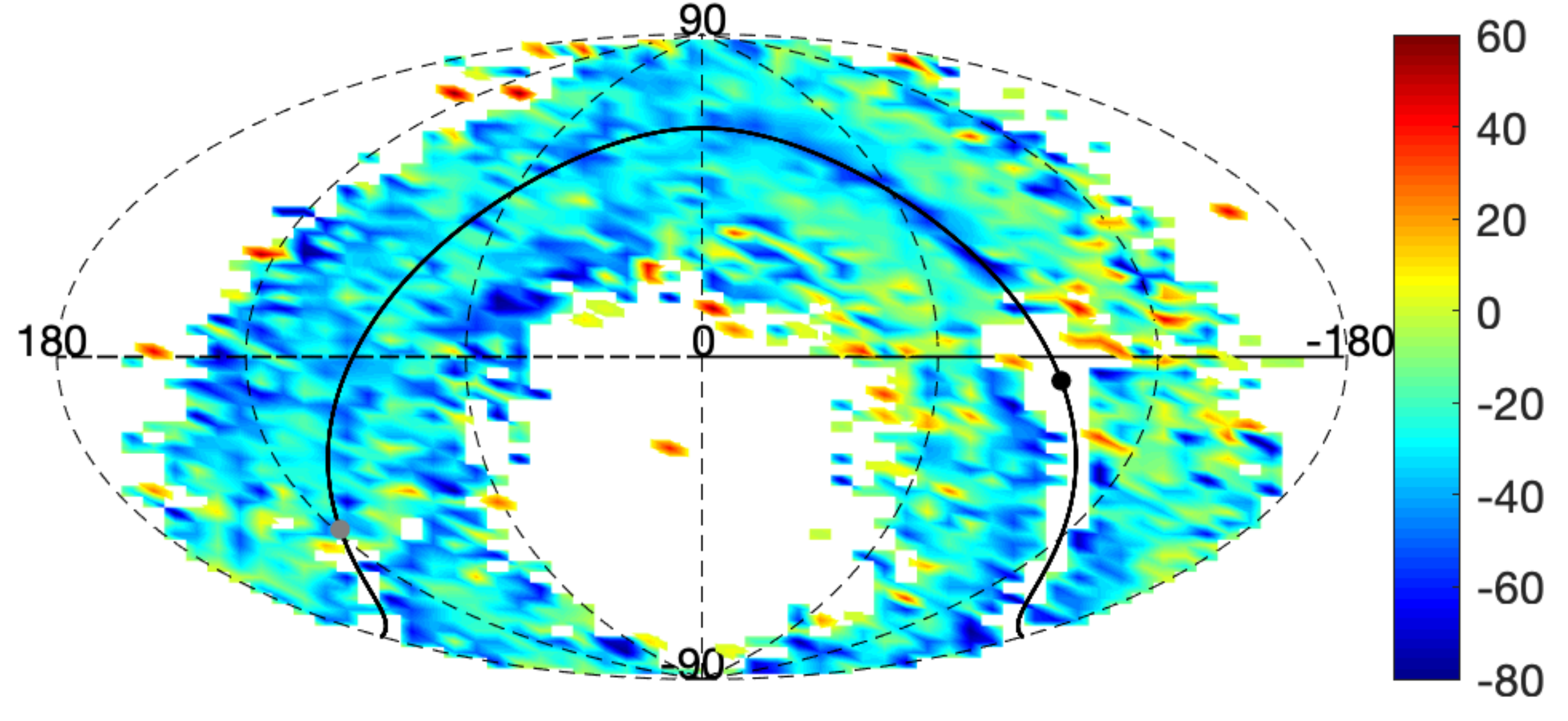}
\includegraphics[width=8cm]{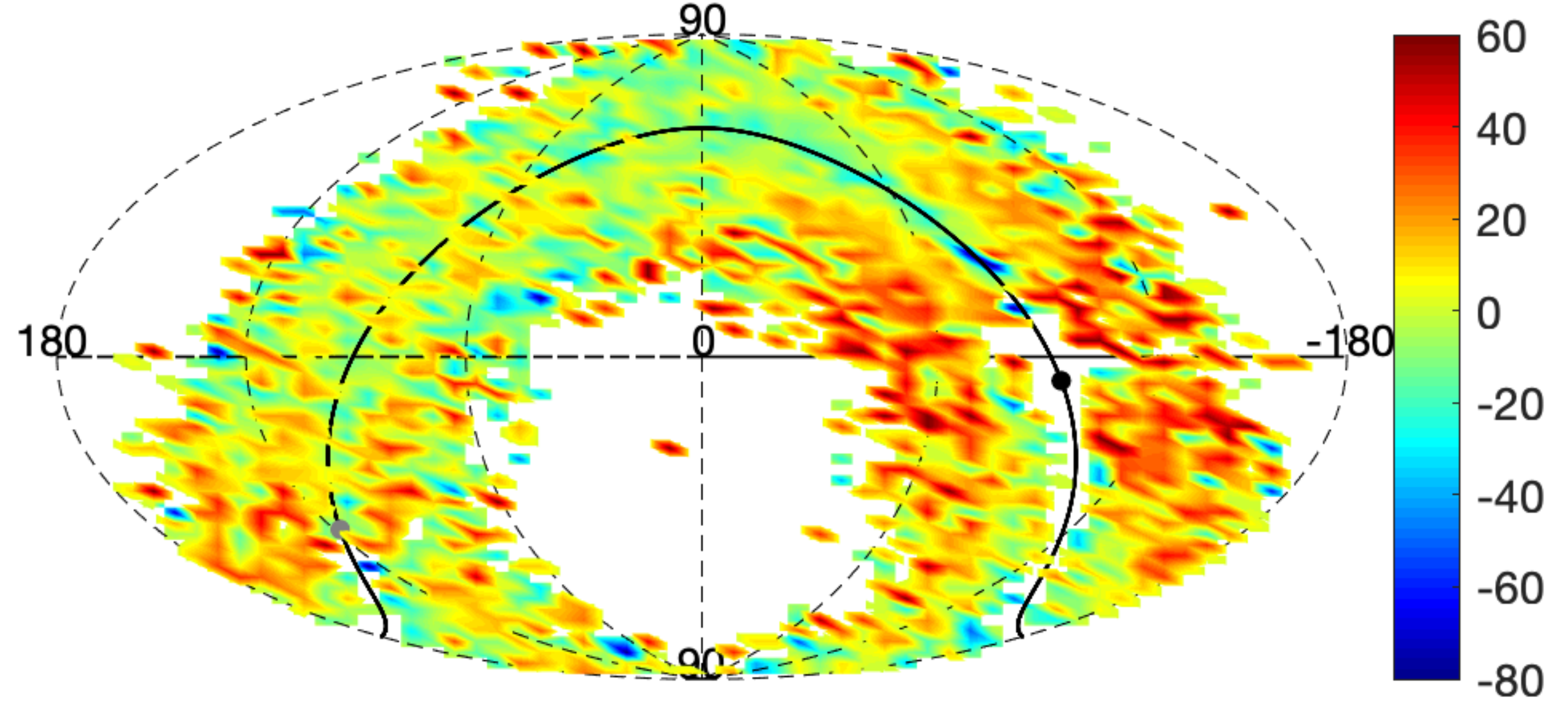}
\includegraphics[width=8cm]{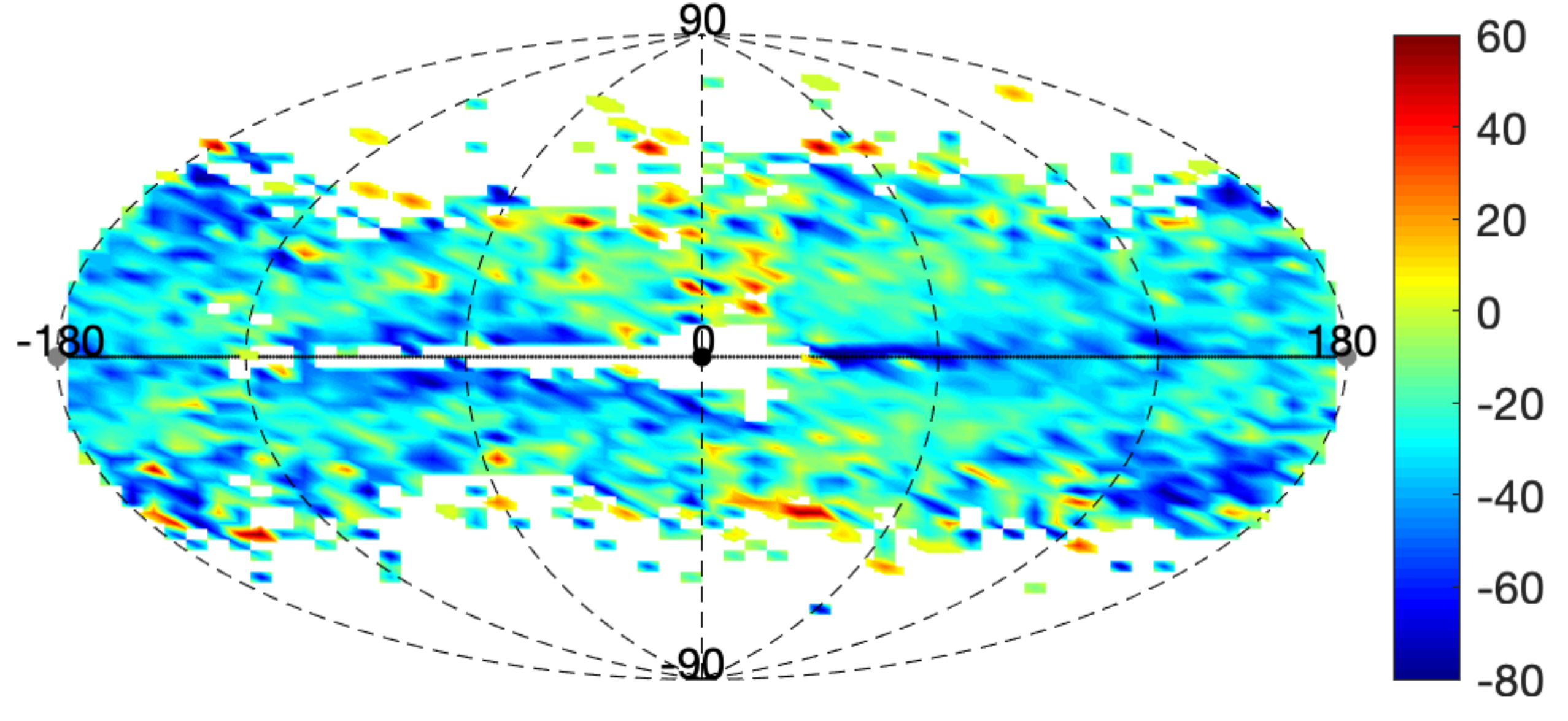}
\includegraphics[width=8cm]{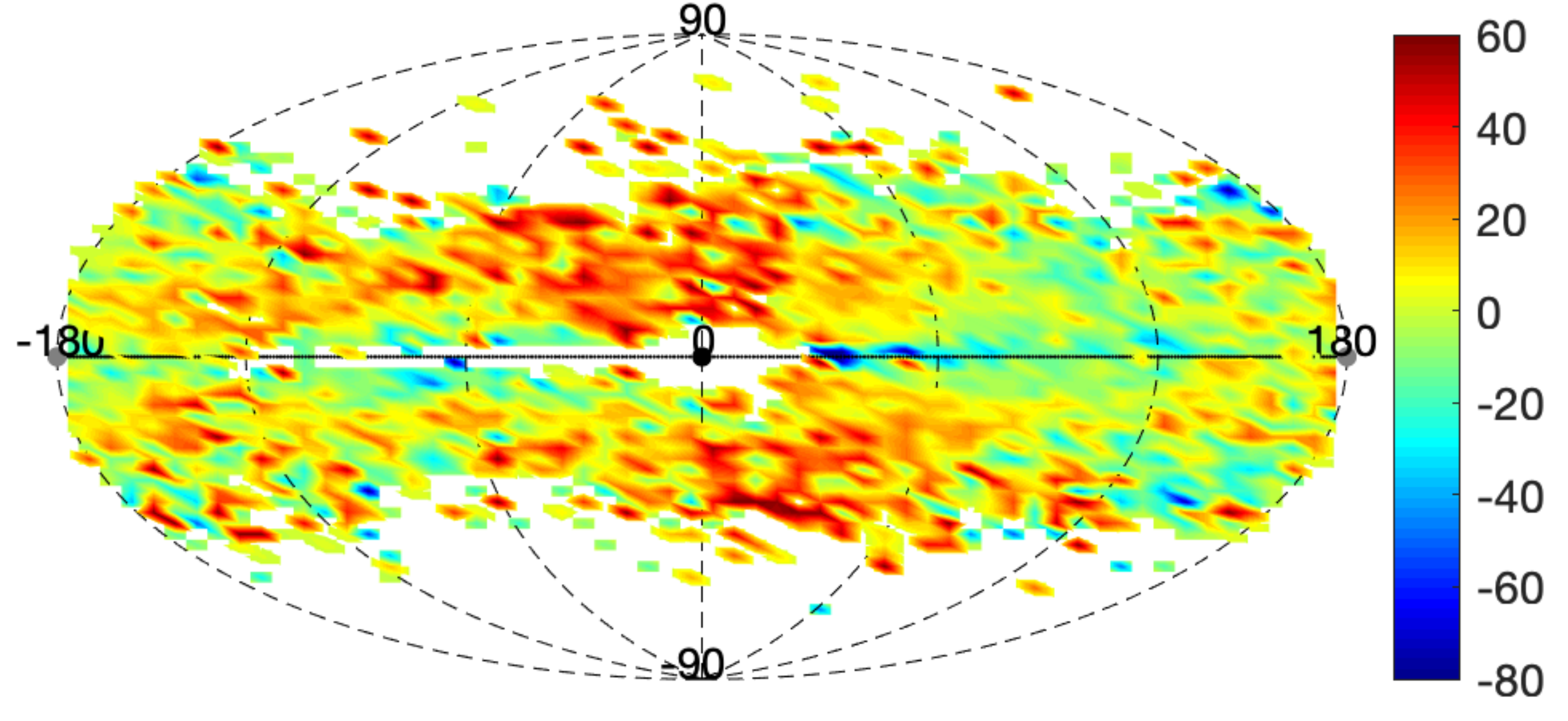}
\caption{Full-sky maps of the mean parallax differences (units:
  $\mu$as; color bars) for the five-parameter solutions of $\Delta
  \pi_{\rm EDR3}$ and $\Delta \pi_{\rm EDR3}^{\rm corr}$. The top two
  panels show Hammer--Aitoff projections of $\Delta \pi_{\rm EDR3}$
  and $\Delta \pi_{\rm EDR3}^{\rm corr}$ in ecliptic coordinates; the
  bottom two panels are in Galactic coordinates. The resolution of
  each pixel before smoothing is 3$\times$3 deg$^2$. We only kept
  pixels with mean parallax differences estimated from more than 10
  EWs. The solid black lines in all panels denote the Galactic
  plane. The Galactic Center and Anti-Center are shown as black and
  gray bullets, respectively.
\label{fig:bias}}
\end{figure}

The $W1$ PLR thus determined can be used to estimate the absolute
magnitudes of all EWs. If the extinction is known, we can obtain an
object's distance. This distance is affected by the PLR zero-point
rather than the {\sl Gaia} parallax zero-point and can therefore be
used to determine the systematic offset in the {\sl Gaia}
parallaxes. For sources covered by the 3D extinction map, we
iteratively obtained the best-fitting extinction, using
$A_{W1}(\mu_0)=m_{W1}-M_{W1}-\mu_0$. For sources not covered by the 3D
extinction map, we estimated the extinction using the $G_{\rm BP}$ and
$G_{\rm RP}$ PLRs, i.e. $A_{W1}=\frac{A_{W1}}{A_{V}} A_V
=\frac{A_{W1}}{A_{V}} \frac{(m_{\rm BP}-M_{\rm BP})-(m_{\rm RP}-M_{\rm
    RP})}{\frac{A_{\rm BP}}{A_{V}}-\frac{A_{\rm RP}}{A_{V}}}$. Here,
$\mu_0$ is the distance modulus, $A_{W1}(\mu_0)$ the extinction at
$\mu_0$, $m_{\lambda}$ and $\rm M_{\lambda}$ are the apparent and
absolute magnitudes in the corresponding band, $\lambda$. We adopted
the \citet{2019ApJ...877..116W} extinction law.

Figure \ref{fig:zero} shows a comparison of the parallaxes derived
from the PLR with those from {\sl Gaia} EDR3. Here, $\pi_{\rm
  EW},\pi_{\rm EDR3}$, and $\pi_{\rm EDR3}^{\rm corr}$ represent
parallaxes obtained from PLR distances, {\sl Gaia} EDR3, and {\sl
  Gaia} EDR3 after zero-point correction based on
\citet{2020arXiv201203380L}, respectively. $\Delta \pi_{\rm
  EDR3}=\pi_{\rm EDR3}-\pi_{\rm EW}$ and $\Delta \pi_{\rm EDR3}^{\rm
  corr}=\pi_{\rm EDR3}^{\rm corr}-\pi_{\rm EW}$ represent parallax
differences.

The parallax differences trace a roughly symmetric, normal
distribution with a negative shift, where the {\sl Gaia} parallaxes
are systematically smaller. The offsets are $\Delta \pi_{\rm
  EDR3}=-28.6\pm0.6$ $\mu$as and $-25.4\pm4.0$ $\mu$as, respectively,
for the five- and six-parameter solutions (Figure \ref{fig:zero}, blue
histograms), where the errors are the standard deviations. These
offsets are slightly larger than that derived from the quasar sample
\citep[$-17$ $\mu$as][]{2020arXiv201201742L}. The $\Delta \pi_{\rm
  EDR3}^{\rm corr}$ distributions are shown as red histograms in
Figure \ref{fig:zero}; the mean values are $4.2\pm0.5$ $\mu$as and
$4.6\pm3.7$ $\mu$as for the five- and six-parameter solutions,
respectively. This suggests that the parallax zero-point correction
provided by the {\sl Gaia} team adopting the quasar reference frame
significantly reduces the bias in the {\sl Gaia} EDR3 parallaxes, but
it may slightly overcorrect the bias for Galactic objects.

Equipped with over 100,000 EWs, we can now assess which parameters
contribute to the systematic offset. Figure \ref{fig:detail} shows the
binned parallax difference distributions, $\Delta \pi_{\rm EDR3}$
(blue dots) and $\Delta \pi_{\rm EDR3}^{\rm corr}$ (red dots), as a
function of $G$ magnitude, effective wavenumber, $\nu_{\rm eff}$,
ecliptic latitude, $\sin\beta$, and Galactic latitude, $\sin b$, for
the five-parameter solutions. Following
\citet{2020arXiv201201742L}, we use $\nu_{\rm eff}$ as a proxy for
the color information. It results from processing of the BP and RP
spectra and can be converted directly to $G_{\rm BP}-G_{\rm RP}$.
The parallax offsets estimated from distant quasars
\citep{2020arXiv201201742L} are shown as gray points for
comparison. Representing one of the largest external comparison
catalogs \citep[e.g.][their Table 1]{2020arXiv201206242F}, the number
of EWs is only smaller than the quasar sample. Meanwhile, EWs have
good coverage in the Galactic plane, spanning a wide range of colors
and magnitudes. The detailed dependence of the parallax differences
derived from EWs is fully complementary to that derived from quasar
analysis; it is more suitable for Galactic stars.

In Figure \ref{fig:detail}a, $\Delta \pi_{\rm EDR3}$ exhibits an
increasing, roughly linear trend as a function of magnitude, similar
to that shown by the quasars at fainter magnitudes. $\Delta \pi_{\rm
  EDR3}^{\rm corr} \simeq 0$ is stable for $14.5<G<17$ mag. Since 55\%
of our EWs are found in this range, we conclude that the offset
analysis based on EWs and quasars is consistent here. In Figure
\ref{fig:detail}b, the $\Delta \pi_{\rm EDR3}$ trends based on EWs and
quasars are consistent for $\nu_{\rm eff}<1.5$ $\mu {\rm m}^{-1}$. For
effective wavenumbers of 1.5--1.64 $\mu$m$^{-1}$, $\Delta \pi_{\rm
  EDR3}$ based on EWs is systematically lower by 10--20 $\mu$as. After
correction, $\Delta \pi_{\rm EDR3}^{\rm corr} \simeq 0$, but the
pattern persists. This effective wavenumber range corresponds to the
range of F- and G-type main-sequence stars. The different offsets may
be associated with different types of stars and this requires
additional data to verify.

We also checked for trends as a function of spatial position. Any
trend in $\Delta \pi_{\rm EDR3}^{\rm corr}$ with ecliptic latitude is
weak (Figure \ref{fig:detail}c). However, a clear trend is seen as a
function of Galactic latitude (Figure \ref{fig:detail}d). Both $\Delta
\pi_{\rm EDR3}$ and $\Delta \pi_{\rm EDR3}^{\rm corr}$ exhibit sharp
drops of 10 $\mu$as in the Galactic plane ($|b| \lesssim 10
\arcdeg$). Elsewhere, $\Delta \pi_{\rm EDR3}$ and $\Delta \pi_{\rm
  EDR3}^{\rm corr}$ exhibit stable distributions around the mean. This
trend is not a result of either extinction or metallicity
variations. It is also present for low-extinction sources. Based on
320 EWs with Large-Area Multi-Object Spectroscopic Telescope (LAMOST)
metallicity measurements, we obtain $\Delta M_{W1} = M_{W1, {\rm obs}}
- M_{W1, {\rm PLR}} = 0.222 {\rm [Fe/H]}+0.014$ mag. The slope agrees
with the near-IR metallicity effect found by
\citet{2016ApJ...832..138C}. If metallicity effects are taken into
account, the trend becomes steeper rather than flatter. Since only
quasars with $|b| > 20 \arcdeg$ are used to model the correction for
the five-parameter solutions, the correction is reliable for disk
sources if the model used for the disk sources is similar to that of
the sources at $|b| > 20 \arcdeg$. However, \citet[][their
  Fig. 13]{2020arXiv201203380L} showed that the rise and fall of the
parallaxes becomes obvious when they approach the Galactic plane. The
likely reason for the trend is, instead, that the correction based on
the five-parameter solutions is insufficient for some regions in the
Galactic disk.

To better investigate the distribution of the parallax differences as
a function of spatial position, maps of $\Delta \pi_{\rm EDR3}$ and
$\Delta \pi_{\rm EDR3}^{\rm corr}$ for the five-parameter solutions,
in both ecliptic and Galactic coordinates, are shown in Figure
\ref{fig:bias}. The parallax correction varies more significantly with
Galactic than ecliptic latitude. The maps are more intuitive to
evaluate the corrected {\sl Gaia} EDR3 parallaxes. After correction,
the parallax offset is less than 10 $\mu$as across 40\% of the sky,
and only 15\% of the sky has a parallax offset greater than 30
$\mu$as. This shows that the correction for {\sl Gaia} EDR3 parallaxes
is effective in reducing deviations in the {\sl Gaia} parallaxes.

\section{Discussion} \label{sec:Discu}

\subsection{Systematic Errors} \label{sec:systematicerror}

Here, we present an estimate of the systematic errors in our PLR-based
EW distances. This is important for assessment as to how accurate our
derived parallax offset is. The systematic uncertainties include four
components, (i) the PLR zero-point offset, (ii) the internal PLR
spread, (iii) unresolved third components, and (iv) errors in our
extinction estimates.

The $W1$ PLR in Section \ref{sec:data} was obtained based on the
parallaxes (calibrated using the five-parameter solution) of 1138
objects located within 500 pc of the Sun. This sample has an average
parallax of 3.06 mas. Taking into account the systematic uncertainty
of $4.2\pm 0.5$ $\mu$as (Figure \ref{fig:zero}), the systematic error
propagating to the PLR contributes 0.15\%. The systematic error
associated with the internal PLR spread is $0.16/\sqrt{1138}=0.0047$
mag, where the $1\sigma$ dispersion of the $W1$ PLR is 0.16 mag. Based
on a study of 75 nearby EWs \citep{2006AJ....132..650D}, the presence
of unresolved third components would affect the parallaxes of our
sample objects by 0.3\%.

For the adopted extinction, the systematic error is contributed by the
extinction difference resulting from application of different methods
of extinction determination and the choice of extinction law. Based on
98,466 EWs, the average extinction difference between the 3D
extinction map used and the extinction calculated from the $G_{\rm
  BP}$ and $G_{\rm RP}$ PLRs is $\Delta A_{W1}=0.0377-0.0394 =
-0.0017$ mag. The average extinction for all EWs is $A_{W1}=0.035$ mag
($A_V=0.895$ mag), which is reliable for an average distance of 2.41
kpc. Considering a 10\% uncertainty in the extinction law, the
systematic bias caused by extinction differences is about
$\sqrt{0.0035^2+0.0017^2}=0.0039$ mag.

We do not consider systematic errors in the PLR caused by metallicity
effects, for two reasons. First, our EWs are distributed uniformly
around the Sun, at an average distance of 2.4 kpc. EW ages are between
1 Gyr and 10 Gyr, and they do not tend to be distributed
preferentially in either the metal-poor halo or the metal-rich
Galactic disk. Therefore, the assumption of an average solar abundance
for the EW PLR as a whole is appropriate. Second, if metallicity
effects are significant, Figure \ref{fig:detail}d shows that the
parallax offset decreases for decreasing $|b|$ (we only consider
$|b|>10\arcdeg$).

Combining the individual error estimates, the systematic uncertainty
affecting our results is $\sigma=415$ $\mu$as $\times$ $[(0.0015)^2+
  (0.0047 \times \ln(10)/5)^2 + (0.003)^2 +(0.0039 \times
  \ln(10)/5)^2]^{1/2}=1.8$ $\mu$as.

\subsection{Comparison}

Recently, much work has been done on {\sl Gaia} EDR3 parallaxes based
on other tracers. \citet{2021ApJ...907L..33S} obtained offsets of
$-37\pm20$ $\mu$as and $-15\pm18$ $\mu$as, respectively, before and
after correction, based on 76 EBS. \citet{2021arXiv210109691H} found a
mean parallax offset of $-26$ $\mu$as based $\sim$70,000 red clump
stars observed with LAMOST, which was reduced to around $4$ $\mu$as
after correction. \citet{2021arXiv210107252Z} and
\citet{2021ApJ...908L...6R} also found an overestimated zero-point
correction of $15\pm3$ $\mu$as and $14\pm6$ $\mu$as based on,
respectively, 2000 first-ascent red-giant-branch stars with
asteroseismic parallaxes in the {\sl Kepler} field and 75 classical
Cepheids. \citet{2021AJ....161..176R} found offsets of
$-42.1\pm1.9\mbox{ (stat.)}\pm12.9\mbox{ (syst.)}\mu$as and
$-10.9\pm2.9\mbox{ (stat.)}\pm12.9\mbox{ (syst.)}\mu$as, respectively,
before and after correction, based on 2334 EWs in the northern
Galactic plane.

Overall, for Galactic stars the {\sl Gaia} EDR3 parallax offsets are
$[-20, -30]$ $\mu$as and 4--10 $\mu$as before and after correction,
respectively. For specific regions---the Galactic disk, the bulge, and
high-latitude regions---there is an additional deviation of about 10
$\mu$as. Compared with previous results, our new results have smaller
errors and higher completeness because of the much larger sample size
afforded by our EW sample and their more complete coverage in
magnitude, color, and spatial distribution.

\section{Conclusion} \label{sec:conclusion}

We have used 109,512 and 4309 EWs with five- and six-parameter
solutions for an independent examination of the {\sl Gaia} EDR3
parallaxes. Our EWs cover the entire sky, except for the Galactic
Center and a small region in the southern Galactic
midplane. Representing one of the largest available catalogs, EW
types have a relatively complete coverage in magnitude and color.

We determined the $W1$ PLR of EWs based on 1194 nearby objects.
Adopting this PLR, we obtained independent parallaxes with a 7.4\%
accuracy to check both the original and zero-point-corrected EDR3
parallaxes. The overall offsets resulting from our analysis are
$\Delta \pi_{\rm EDR3}=-28.6\pm0.6$ $\mu$as and $\Delta \pi_{\rm
  EDR3}^{\rm corr}=4.2\pm0.5$ $\mu$as for five-parameter solutions,
and $\Delta \pi_{\rm EDR3}=-25.4\pm4.0$ $\mu$as and $\Delta \pi_{\rm
  EDR3}^{\rm corr}=4.6\pm3.7$ $\mu$as for six-parameter solutions,
with a systematic uncertainty of 1.8 $\mu$as. The relationships, if
any, between the parallax offset and $G$-band magnitude, effective
wavenumber, $\nu_{\rm eff}$, ecliptic latitude, $\sin\beta$, and
Galactic latitude, $\sin b$, were investigated. The EW results
generally agree with those derived from quasars, except for the
smaller parallax offsets for effective wavenumbers of 1.5--1.64 $\mu
m^{-1}$ and for stars in the Galactic plane ($|b| \lesssim 10
\arcdeg$). We found that any correlation between parallax offsets and
ecliptic latitude is weak.

The spatial distribution of the parallax offsets shows that the bias
in corrected {\sl Gaia} EDR3 parallaxes is less than 10 $\mu$as across
40\% of the sky. Only 15\% of the sky is affected by parallax offsets
greater than 30 $\mu$as. We have thus provided independent evidence
that the {\sl Gaia} EDR3 parallax corrections are effective. Combined
with literature data, we found that the overall offsets in {\sl Gaia}
EDR3 parallaxes for Galactic stars are $[-20, -30]$ $\mu$as and 4--10
$\mu$as, respectively, before and after correction. For specific
regions, such as the Galactic disk, the bulge, and high-latitude
regions, there is an additional deviation of about 10
$\mu$as. Compared with {\sl Gaia} DR2, the parallax accuracy of {\sl
  Gaia} EDR3 is thus greatly improved.

\acknowledgments 
We are grateful for research support from the National Key Research
and Development Program of China through grants 2019YFA0405500 and
2017YFA0402702. We also received support from the National Natural
Science Foundation of China through grants 11903045 and 11973001. This
work has made use of data from the European Space Agency' {\sl Gaia}
mission \\(http://www.cosmos.esa.int/gaia), processed by the {\sl
  Gaia} Data Processing and Analysis Consortium (DPAC;
\\ http://www.cosmos.esa.int/web/gaia/dpac/consortium).

\bibliographystyle{aasjournal_V1}
\bibliography{EB_EDR3}

\end{document}